\documentclass[aps,prb,superscriptaddress,byrevtex,preprint]{revtex4-1}

\pdfoutput=1
\usepackage{color}
\usepackage[pdftex]{graphicx}
\usepackage{epstopdf}
\usepackage[latin1]{inputenc}
\usepackage[ngerman,english]{babel}

\begin{document}

\title{The importance of charge fluctuations for the topological phase in SmB$_6$}
\date{\today}

\author{Chul-Hee Min}
\email[corresponding author. Email:]{cmin@physik.uni-wuerzburg.de}
\affiliation{Universit\"at W\"urzburg, Experimentelle Physik VII \& R\"ontgen Research Center for Complex Material Systems RCCM, 97074 W\"urzburg, Germany}
\affiliation{Karlsruhe Institut f\"ur Technologie KIT, Gemeinschaftslabor f\"ur Nanoanalytik, 76021 Karlsruhe, Germany}
\author{P. Lutz}
\affiliation{Universit\"at W\"urzburg, Experimentelle Physik VII \& R\"ontgen Research Center for Complex Material Systems RCCM, 97074 W\"urzburg, Germany}
\affiliation{Karlsruhe Institut f\"ur Technologie KIT, Gemeinschaftslabor f\"ur Nanoanalytik, 76021 Karlsruhe, Germany}
\author{S. Fiedler}
\affiliation{Universit\"at W\"urzburg, Experimentelle Physik VII \& R\"ontgen Research Center for Complex Material Systems RCCM, 97074 W\"urzburg, Germany}
\affiliation{Karlsruhe Institut f\"ur Technologie KIT, Gemeinschaftslabor f\"ur Nanoanalytik, 76021 Karlsruhe, Germany}

\author{B. Y. Kang}
\affiliation{School of Materials Science and Engineering, Gwangju Institute of Science and Technology (GIST), Gwangju 500-712, Korea.}
\author{B. K. Cho}
\email[corresponding author about single crystals and basic characterization. Email:]{chobk@gist.ac.kr}
\affiliation{School of Materials Science and Engineering, Gwangju Institute of Science and Technology (GIST), Gwangju 500-712, Korea.}
\author{H. -D. Kim}
\affiliation{Center for Correlated Electron Systems, Institute for Basic Science (IBS), Seoul 151-747, Republic of Korea}
\affiliation{Department of Physics and Astronomy, Seoul National University, Seoul 151-747, Republic of Korea}

\author{H. Bentmann}
\affiliation{Universit\"at W\"urzburg, Experimentelle Physik VII \& R\"ontgen Research Center for Complex Material Systems RCCM, 97074 W\"urzburg, Germany}
\affiliation{Karlsruhe Institut f\"ur Technologie KIT, Gemeinschaftslabor f\"ur Nanoanalytik, 76021 Karlsruhe, Germany}
\author{F. Reinert}
\affiliation{Universit\"at W\"urzburg, Experimentelle Physik VII \& R\"ontgen Research Center for Complex Material Systems RCCM, 97074 W\"urzburg, Germany}
\affiliation{Karlsruhe Institut f\"ur Technologie KIT, Gemeinschaftslabor f\"ur Nanoanalytik, 76021 Karlsruhe, Germany}

\maketitle 
\textbf{The discovery of topologically non-trivial states in band insulators \cite{fu_topological_2007-1,hasan_colloquium:_2010} has induced an extensive search for topological phases in strongly correlated electron systems \cite{dzero_topological_2010}. In particular, samarium hexaboride (SmB$_6$) has drawn a lot of attention as it represents a new class of condensed matter called topological Kondo insulator \cite{dzero_topological_2010,wolgast_discovery_2012}. Kondo insulators (KI) \cite{riseborough_heavy_2000,nozawa_ultrahigh-resolution_2002,souma_direct_2002} can have non-trivial Z$_2$ topology \cite{dzero_topological_2010} because the energy gap opens at the Fermi energy ($E_F$) by a hybridization between a renormalized $f$ band with odd parity and a  conduction $d$ band with even parity \cite{takimoto_smb<sub>6</sub>:_2011,alexandrov_cubic_2013}. However, SmB$_6$ deviates from the conventional KI \cite{pietrus_kondo-hole_2008} because its gap is insensitive to doping and pressure \cite{cooley_smb_6:_1995, falicov_valence_1981,martin_theory_1979,varma_mixed-valence_1976}. Thus, it is unclear what makes the gap of SmB$_6$ different from that of KI, and how the band inversion occurs. Here we demonstrate the importance of charge fluctuations in SmB$_6$ for the existence of topological phases. Our angle-resolved photoemission spectroscopy (ARPES) results reveal that with decreasing temperature the bottom of the \,$d$-$f$\,hybridized band at the $\bar{\text{X}}$\,point gradually shifts from below to above $E_F$, accompanied by a characteristic redistribution of spectral weight. Moreover, because this hybridized band is predicted to have odd parity and to induce a non-trivial $Z_{2}$ topology \cite{takimoto_smb<sub>6</sub>:_2011, lu_correlated_2013,alexandrov_cubic_2013}, we investigated the existence of topological surface states. Our results show that SmB$_6$ is a correlated topological insulator in which charge fluctuations \cite{martin_theory_1979,varma_mixed-valence_1976}do not only drive the unusual metal-insulator transition but are responsible for the existence of a non-trivial topological phase.} 

\indent The constant resistivity of bulk-insulating SmB$_6$ at $T<5$\,K (Supplementary Fig.\,S1) has been a long-standing puzzle because such a behaviour cannot be explained by the classical contributions from impurities \cite{allen_large_1979}. After ``topology'' has been found to be a fundamental concept for the description of the electronic properties of solids \cite{fu_topological_2007-1,hasan_colloquium:_2010}, specially including strongly correlated systems \cite{dzero_topological_2010}, the region of the constant resistivity has been interpreted as the signature of a non-trivial topological phase with topologically protected surface states \cite{wolgast_discovery_2012,kim_limit_2012,yeo_effects_2012}. Despite extensive experimental and theoretical investigations on SmB$_6$, the electronic structure and its connection to the temperature dependence are not fully understood up to now --- mainly because of the complex interplay among strong electron correlations, spin-orbit interactions, and multiplet splitting. We present here a detailed experimental approach by high-resolution ARPES with the focus on the temperature dependence of surface and bulk electronic structure.\\
\begin{figure}
	\centering
			\includegraphics[width=1.0\textwidth]{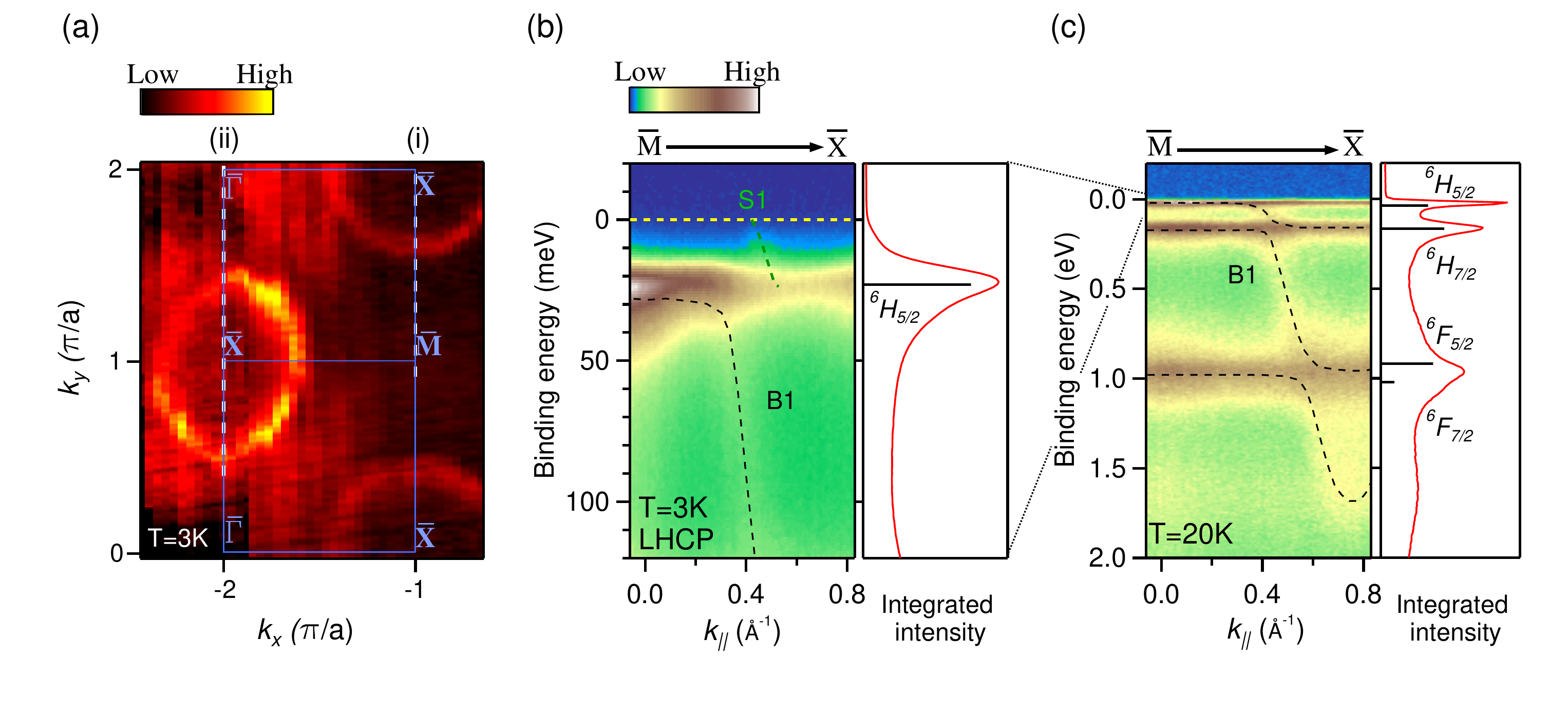}
	\caption{\textbf{Overview of the electronic structure of SmB$_6$.} \textbf{(a)} Fermi surface (FS) map of SmB$_6$ obtained at $T$\,$=$\,3\,K using $h\nu$\,$=$\,70\,eV. Elliptical FS sheets appear around the $\bar{\text{X}}$\,points. ARPES spectra were obtained along the white lines (i) and (ii) representing the high symmetry directions $\bar{\text{\rm M}}$--$\bar{\text{X}}$, and $\bar{\text{X}}$--$\bar{\Gamma}$. \textbf{(b),\,left:} ARPES intensity vs.~binding energy ($E_B$) and momentum ($k_\parallel$) along (i). The surface state S1 \cite{xu_surface_2013, jiang_observation_2013} indicated by the green dashed line crosses $E_F$ ($E_B=0$). The bulk-\,quasiparticle band B1 is indicated by black dashed lines. \textbf{(b),\,right:} integrated Energy Distribution Curve (EDC) from the data on the left. The $^{6}H_{5/2}$\,final state multiplet is positioned at $E_B$\,$=$\,21\,meV. \textbf{(c),\,left:} wide energy range ARPES spectrum measured along (i). As illustrated with black dashed lines, B1 does not show a simple parabolic dispersion \cite{xu_surface_2013, jiang_observation_2013} because its slope changes due to hybridization with the $^{6}H_{5/2}$, $^{6}H_{7/2}$, and $^{6}F$\,final state multiplets for the $f^6$\,$\rightarrow$\,$f^5$\,transitions. \textbf{(c),\,right:} integrated EDC from the left with calculated multiplet lines \cite{gerken_calculated_1983}.}
	\label{fig:fig1}
\end{figure}
\indent Fig.\,\ref{fig:fig1} shows the general electronic structure of SmB$_6$: (a) on the left gives the surface Brillouin zone (SBZ) of SmB$_6$ with the elliptical Fermi surface (FS) sheets around the $\bar{\text{X}}$\,points. The details of the electronic structure are analyzed along the white dashed lines (i) and (ii), corresponding to the high symmetry lines $\bar{\rm M}$--$\bar{\text{X}}$ and $\bar{\text{X}}$--$\bar{\Gamma}$, respectively. Fig.\,\ref{fig:fig1}\,(b) shows the ARPES intensity as a function of binding energy $E_B$ ($y$-axis) and momentum $k_\parallel$ ($x$-axis) along $\bar{\rm M}$--$\bar{\text{X}}$ at low temperatures, i.e. at $T$\,$=$\,3\,K. The surface state S1 (green dashed line) \cite{xu_surface_2013, jiang_observation_2013} crosses $E_F$ and forms the observed elliptic sheets in the FS plots in (a). The Sm\,$4f$ states appear in the  photoemission data as final-state multiplets for the $f^6$\,$\rightarrow$\,$f^5$\,transition. The right sides of Fig.\,\ref{fig:fig1}\,(b)\,$\&$\,(c) show the integrated energy distribution curves (EDC) together with black bars, which denote the calculated relative intensities and energy positions of the $^{6}H_{5/2}$, $^{6}H_{7/2}$, and $^{6}F$ contributions \cite{chazalviel_study_1976,gerken_calculated_1983}. In the angle resolved data, the multiplets appear as narrow bands that hybridize with the bulk $d$\,band B1 \cite{xu_surface_2013, jiang_observation_2013}, clearly visible by a changing slope of B1 within the three displayed energy regions. Such a hybridization is only possible if the multiplets have the same symmetry as B1, leading to hybridization gaps between B1 and the narrow $4f$\,bands \cite{martin_theory_1979}.\\
\begin{figure}
  \centering
			\includegraphics[width=0.93\textwidth]{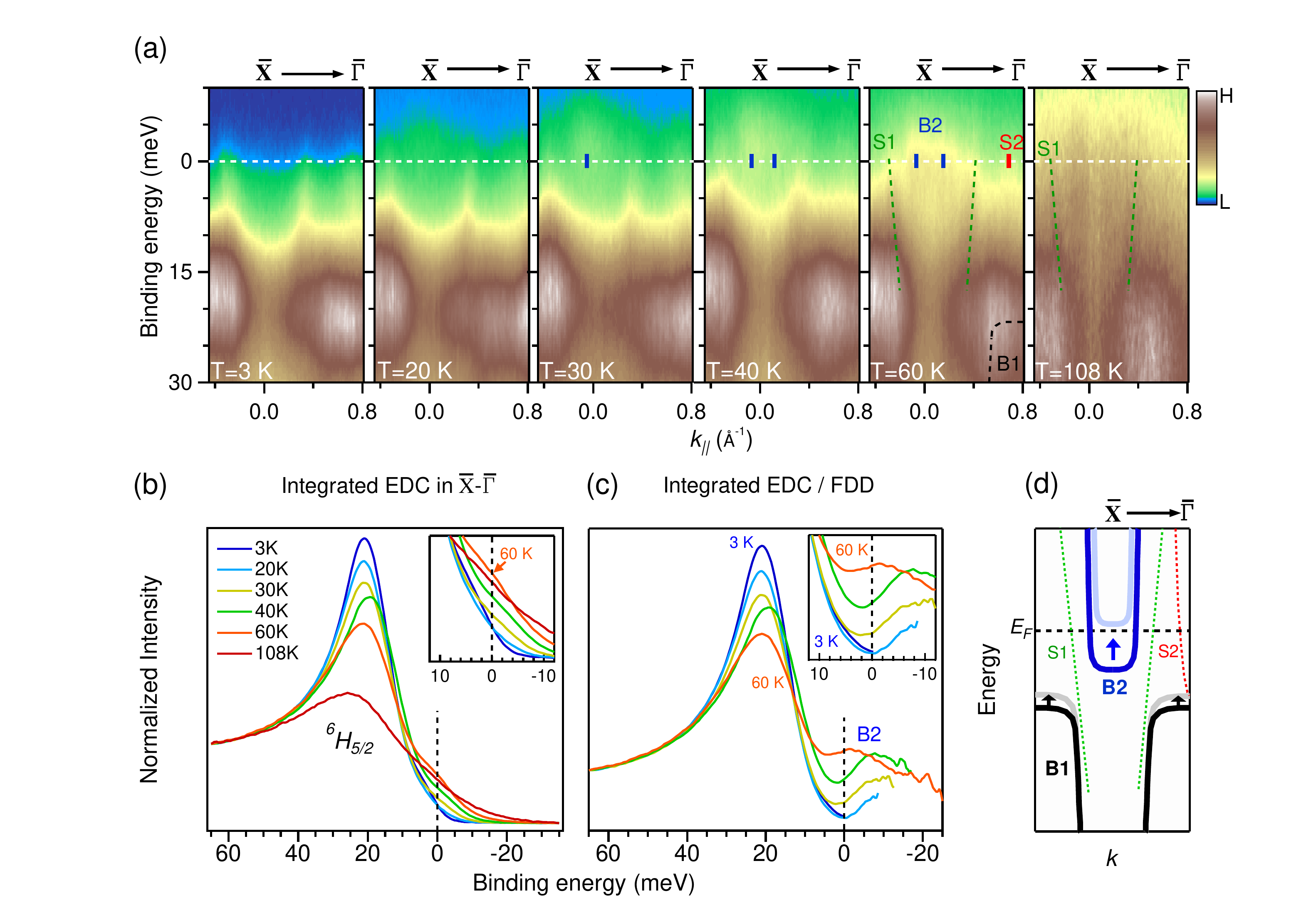}
	\caption{\textbf{Gap evolution in the $\bar{\text{X}}$-$\bar{\Gamma}$ direction.} \textbf{(a)} Temperature-dependence of the ARPES data (vs.~$E_B$ and $k_\parallel$). A lower and an upper quasiparticle band B1 \cite{xu_surface_2013, jiang_observation_2013} and B2 are indicated in black and blue; the surface states S1 and S2 \cite{xu_surface_2013, jiang_observation_2013} are indicated in green and red, respectively. At the $\bar{\text{X}}$\,point, B2 appears below $E_F$ at $T$\,$\geq$\,60\,K, and gradually vanishes with decreasing temperature. \textbf{(b)} Temperature-dependent integrated EDC between the $\bar{\text{X}}$ and $\bar{\Gamma}$\,point in (a). The $^{6}H_{5/2}$ final state multiplet develops with decreasing temperature. \textbf{Inset in (b)} The EDC at $T$\,$=$\,60\,K has the highest spectral weight at $E_F$. \textbf{(c)} EDC for \,3\,K\,$\leq$\,$T$\,$\leq$\,60\,K divided by the Fermi-Dirac distribution (FDD). B2 shifts from below to above $E_F$. \textbf{Inset in (c)} Enlarged spectra near $E_F$. At $T$\,$=$\,60\,K $E_F$ is positioned at the bottom of B2, and with further decreasing temperature it shifts to the middle of the gap. \textbf{(d)} Schematic energy-\,band diagram illustrating the observed band structure in (a). The energy shifts of B1 and B2 with decreasing temperature are indicated by arrows.}
 \label{fig:fig2}
\end{figure}
\indent In order to investigate the temperature dependent development of the gap, we performed ARPES measurements along $\bar{\text{X}}$--$\bar{\Gamma}$ (Fig.\,\ref{fig:fig2}\,(a)) for various temperatures. At $T$\,$=$\,60\,K --- i.e. at an intermediate temperature --- one can observe a lower (B1, marked black) and an upper quasiparticle band (B2, blue). At this temperature, the upper band B2 appears around $\bar{X}$ with a maximum binding energy slightly below $E_F$, but it gradually vanishes with decreasing temperature. As a consequence, the density of states at $E_F$ decreases, as shown in the angle-integrated EDC in Fig.\,\ref{fig:fig2}\,(b). In addition, one can observe that the dominating peak of the $^{6}H_{5/2}$ multiplet increases in intensity and decreases in width with decreasing temperature \cite{nozawa_ultrahigh-resolution_2002, souma_direct_2002}. The narrowing of the 4$f$\,features results in well-developed quasiparticle bands and thus indicates that the temperature approaches the coherence temperature (Supplementary Fig.\,S2). In order to investigate the spectra near $E_F$ up to an energy of $\approx 5k_BT$ above $E_F$, we divided the integrated spectra by the Fermi-Dirac distribution (Fig.\,\ref{fig:fig2}\,(c)) \cite{ehm_quantitative_2002}. Through this well-established normalization, we are able to see that B2, leading to a characteristic intensity slightly above $E_F$, shifts from below to above $E_F$ with decreasing temperature. On the other hand, the intensity of the strong $^{6}H_{5/2}$ peak increases, which implies that the spectral weight of B2 is mainly redistributed to this $4f$\,feature when it disappears from the gap region at $E_F$. The temperature dependence of the gap region is shown in the inset of Fig.\,\ref{fig:fig2}\,(c), in detail: at $T$\,$=$\,60\,K the gap is filled because B2 crosses $E_F$, at 40\,K the gap opens, and at low temperatures $E_F$ shifts towards the middle of the gap. The schematic diagram in Fig.\,\ref{fig:fig2}\,(d) illustrates the observed band structure in (a) and the energy shifts of B1 and B2 with respect to $E_F$.\\
\begin{figure}
	\centering
				\includegraphics[width=1.0\textwidth]{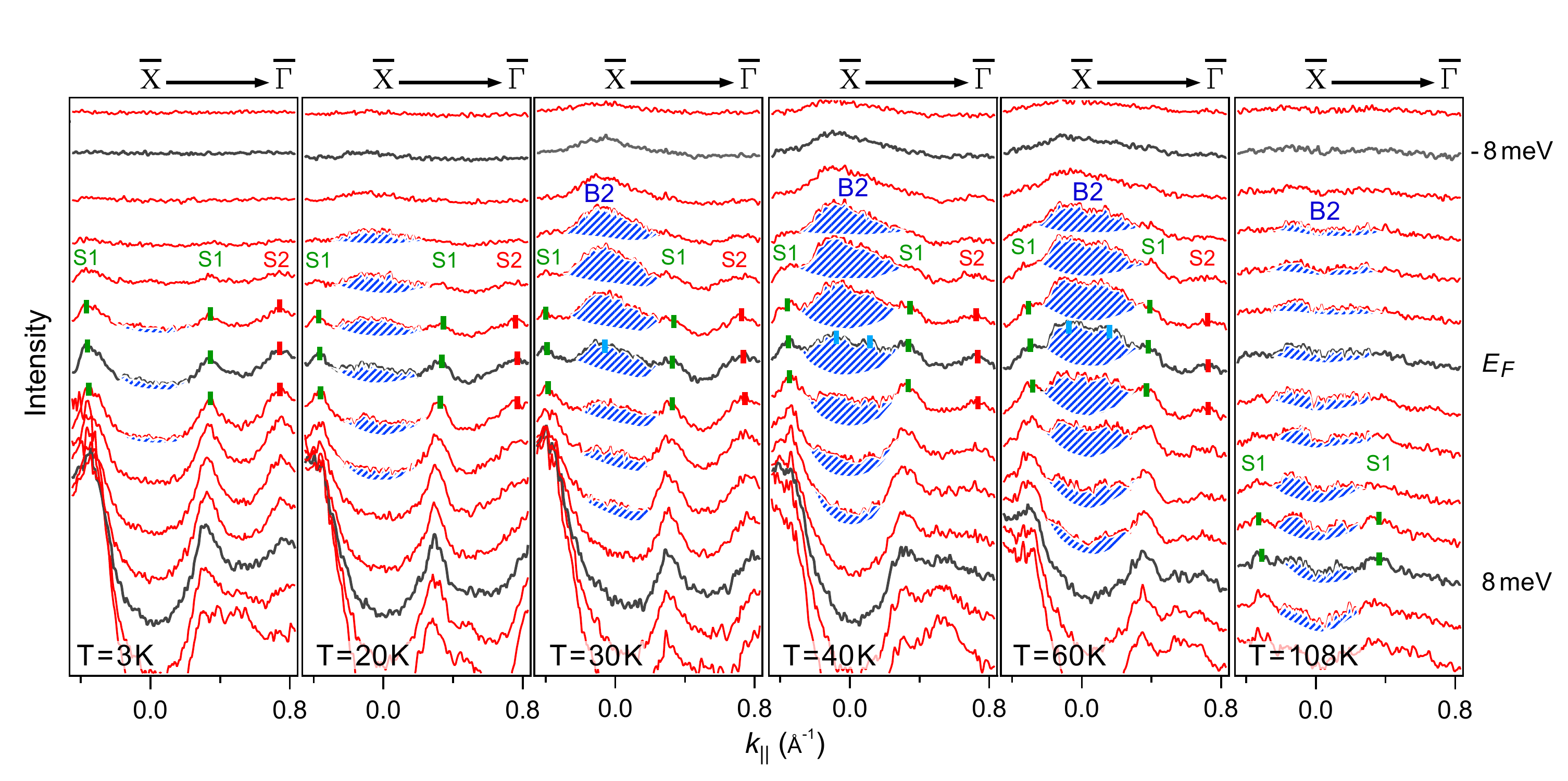}
	\caption{\textbf{Hybridized band B2 shifting above the Fermi level.} Momentum distribution curves (MDC) taken from the data in Fig.\,\ref{fig:fig2}\,(a) close to $E_F$. Bars indicate the peak positions for B2, S1, and S2 with the specific colors. The blue dashed areas denote the spectral weight of B2. The spectral weight of B2 close to $E_F$ reduces for $T$\,$<$\,60\,K. The intensity of B2 becomes narrower with decreasing temperature indicating that B2 gradually shifts upwards.}
	\label{fig:fig3}
\end{figure}
\indent To investigate the behaviour of B2 in more detail, we took momentum distribution curves (MDC) from the ARPES data in the range $\mid$$E_B$$\mid$\,$\leq$\,8\,meV (Fig.\,\ref{fig:fig3}). Between the two S1 peaks (green bars), considerable spectral weight of B2 is present near $E_F$, indicated by the blue shaded area. The area is drawn as a guide to the eye using U-shaped background lines extracted from MDC at higher $E_B$. With decreasing temperature below 60\,K, the spectral weight of B2 below $E_F$ is reduced. Moreover, the two peaks due to B2 at $E_F$ (cyan bars) approach each other. This indicates that B2 gradually shifts above $E_F$ as sketched in Fig.\,\ref{fig:fig2}\,(d). Therefore, the gap evolution of SmB$_6$ deviates from that of a conventional KI whose gap starts to open right at $E_F$. At $T$\,=\,3\,K, there remains small finite weight that possibly originates from the incoherent part of B2 or from the predicted topological surface states (TSS) \cite{takimoto_smb<sub>6</sub>:_2011, lu_correlated_2013, alexandrov_cubic_2013}.\\
\indent The overall band structure near $E_F$ observed in our ARPES spectra consisting of the lower and upper $d$-$f$ hybridized bands B1 and B2 is in agreement with the predictions from {\em ab initio} and model-based theoretical approaches \cite{takimoto_smb<sub>6</sub>:_2011, lu_correlated_2013, alexandrov_cubic_2013}. In addition, the proposed scheme in Fig.\,\ref{fig:fig2}\,(d) is further substantiated by the reduction of spectral weight of the $4f$\,band ($^{6}H_{5/2}$) around the $\bar{\text{X}}$\,point (Fig.\,S2). Therefore, we consider B2 as a bulk derived feature. The determined gap evolution is in agreement with crucial features in the physical properties of SmB$_6$ above 30\,K. The steep slope in resistivity at $T$\,$=$\,40\,K (Fig.\,S1), the maximum number of electron carriers at $T$\,$=$57\,K detected in Hall measurements \cite{nickerson_physical_1971, allen_large_1979}, and the broad peak in the heat capacity at $T$\,$\approx$\,40\,K \cite{nefedova_thermodynamic_2003} can be understood as the consequence of the shift of the bulk band B2, which leads to the unusual (semi)metal-insulator phase transition \cite{nickerson_physical_1971}. Furthermore, the shift can explain the significant change in $f$\,electron density occurring at $T$\,$>$\,30\,K in the X-ray absorption data \cite{mizumaki_temperature_2009}.\\ 
\indent Since the $4f$\,states are located near $E_F$ in mixed valence systems \cite{martin_theory_1979,varma_mixed-valence_1976} i.e. SmB$_6$, thermally excited charge fluctuations between localized $f$ and itinerant $d$\,states determine the thermodynamic properties \cite{hewson__1993}. The influence of charge fluctuations on the electronic structure near $E_F$ is observed in our data by the shift of B2 and the spectral weight redistribution. The shift reflects the change in the carrier density \cite{allen_large_1979} and the increase in the localized $f$\,electron density \cite{mizumaki_temperature_2009}. Thus, our results give a microscopic explanation for the unique temperature dependence of the physical properties of SmB$_6$, and indicate clearly that charge fluctuations play the major role in the phase transition of SmB$_6$ so that it belongs at $T$\,$>$\,30\,K to the charge fluctuation regime of the Anderson model \cite{coleman_introduction_2011} (Supplementary Fig.\,S3).\\ 
\indent Another important aspect follows from the fact that B2 is expected to have odd parity at the $\rm X$\,point and lies above $E_F$ in the ground state. This means that SmB$_6$ is in a strong topological insulator phase with odd Z$_2$ topological invariants \cite{takimoto_smb<sub>6</sub>:_2011,lu_correlated_2013,alexandrov_cubic_2013}. Therefore, we investigated the electronic structure for signatures of a non-trivial topological phase, which can be proven by the existence of TSS. In fact, we observe three different surface states: S1 at the $\bar{\text{X}}$\,point \cite{xu_surface_2013, jiang_observation_2013,neupane_surface_2013}, S2 at the \,$\bar{\Gamma}$\,point \cite{xu_surface_2013, jiang_observation_2013}, and the new state S3 also at the $\bar{\text{X}}$\,point. The surface states S1 and S2 indicated in green and red, respectively (Figs.\,\ref{fig:fig2}\,$\&$\,\ref{fig:fig3}), are located near $E_F$. S2 has been theoretically proposed to be TSS \cite{takimoto_smb<sub>6</sub>:_2011, lu_correlated_2013, alexandrov_cubic_2013} whereas S1 has not been predicted yet. S1 might be a surface band originating from the Sm ions at the surface having a different valence from that in the bulk \cite{aono_lab6_1979}.\\ 
\begin{figure}
	\centering
				\includegraphics[width=1.0\textwidth]{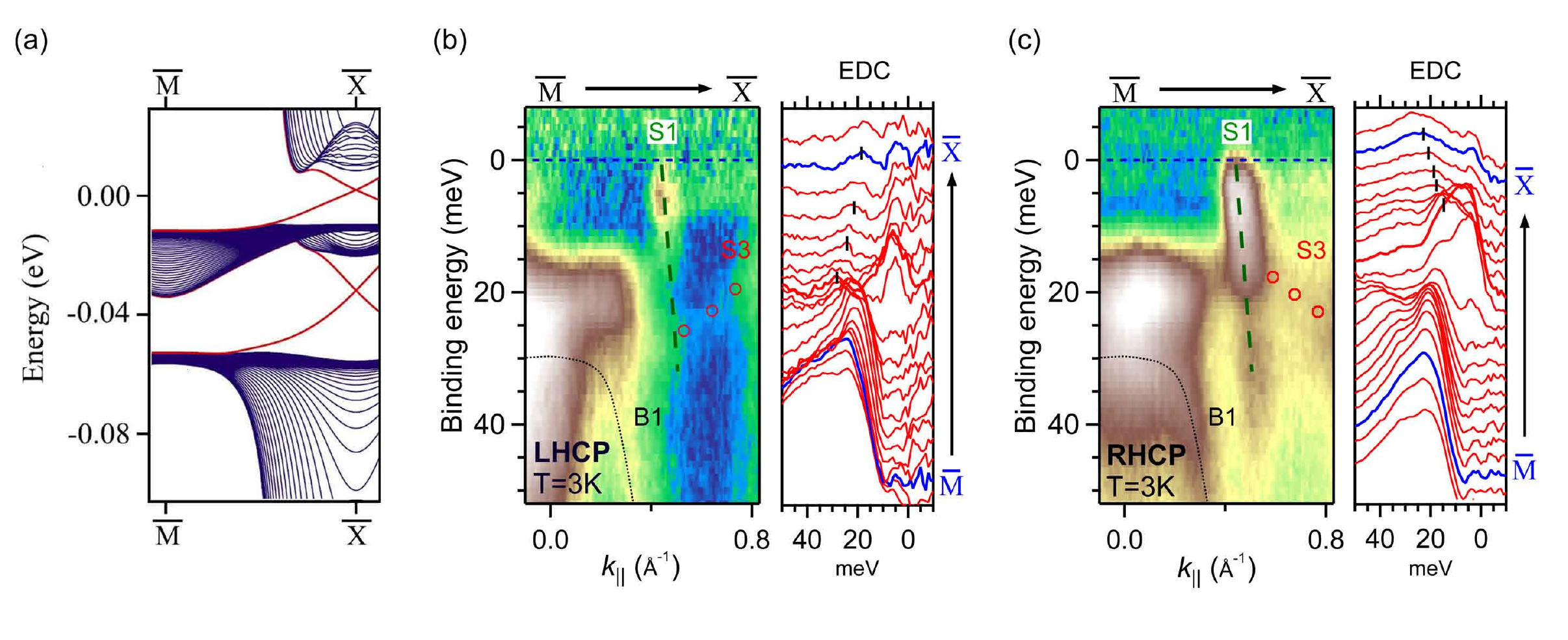}
	\caption{\textbf{Band structure along $\bar{\text{M}}$--$\bar{\text{X}}$.} \textbf{(a)}  Calculated projected bulk band structures and surface states on the (001) surface from Ref.\,\onlinecite{lu_correlated_2013}. The blue and red lines are the projected bulk bands and the topological surface states, respectively. \textbf{(b)\,$\&$\,(c),\,left:} ARPES intensity maps along $\bar{\text{M}}$-$\bar{\text{X}}$ on a logarithmic intensity scale. Depending on the helicity of the circularly polarized light, one branch (red circles) of surface state S3 is selectively enhanced in intensity. Compared to the theoretical calculation, bands S3 could be topological surface states. \textbf{(b)\,$\&$\,(c),\,right:} respective EDC cuts from the data on the left.}
	\label{fig:fig4}
\end{figure}
\indent In addition to the states S1 and S2, our data show the existence of a Dirac cone like band S3 at $E_B$\,=\,20\,meV and $\bar{\text{X}}$, as predicted by the band calculations \cite{lu_correlated_2013} (Fig.\ref{fig:fig4}\,(a)). The red and blue lines represent TSS and bulk bands projected on the (001) surface, respectively. In order to separate dispersive bands from $k$-independent features \cite{martin_theory_1979} in Fig.\,\ref{fig:fig4}\,(b)\,$\&$\,(c), the ARPES intensity is plotted on a normalized logarithmic scale along $\bar{\text{M}}$-$\bar{\text{X}}$, and compared with the theoretical band structure. The calculated band structure has to be shrunk in energy to match B1 hybridizing with the $^{6}H_{5/2}$ feature (Supplementary Fig.\,S4). By circular dichroism in the photoemission process (Fig.\,\ref{fig:fig4}\,(b)\,$\&$\,(c), left), one of the two bands (red circles) crossing at the $\bar{\text{X}}$\,point is separately enhanced, which is a typical dichroism behavior for spin-helical Dirac cones \cite{wang_circular_2013}. These bands S3 are in reasonable agreement with calculated surface states (red lines in Fig.\,\ref{fig:fig4}\,(a)). The corresponding EDC (Fig.\,\ref{fig:fig4}\,(b)\,$\&$\,(c), right) further confirm the existence of the Dirac cone like bands, which appear as dispersing peaks marked by black lines.\\
\indent The comparison of the theoretical and experimental band structure given in Figs.\,\ref{fig:fig4}\,$\&$\,S4 confirms that bands S3 disperse as the predicted TSS that appear within the local gap region of the projected bulk band structure --- in addition to S2 \cite{xu_surface_2013, jiang_observation_2013}. Thus, SmB$_6$ can be in the strong topological phase as predicted \cite{takimoto_smb<sub>6</sub>:_2011, lu_correlated_2013, alexandrov_cubic_2013} because the odd-\,parity band B2 is positioned above $E_F$ at the $X$\,point corroborated by our results. As has been demonstrated, the states at $E_F$ are strongly influenced by charge fluctuations, which drive the position of $E_F$ into the middle of the gap and are responsible for the band inversion that eventually makes SmB$_6$ different from a conventional KI.\\
\section{Methods summary}
 High-quality SmB$_6$ single crystals were cleaved $in$-$situ$ along the (001) plane. The high-resolution ARPES measurements were carried out at the UE112-PGM-1b (``$1^{3}$'') beamline of BESSY~II using $h\nu$\,=\,70\,eV at 3\,K\,$\leq$\,$T$\,$\leq$\,108\,K at Helmholtz-Zentrum Berlin (HZB). The obtained energy resolutions were 7\,meV and 10\,meV for detailed data and the Fermi surface maps, respectively. The accuracy in determining the high symmetry lines in $k$-space was $\pm$\,1$^{\circ}$. (Full method is in Supplementary).\\

\section{acknowledgments}
We thank HZB for the allocation of synchrotron radiation beamtime and the financial support. In particular, we would like to thank Dr. Emile Rienks for assistance during the beamtime, and EP7 colleagues for supports, specially F. Boariu, and former NSMML members for the characterization, particularly J. Y. Kim, and S. S. Lee. We thankfully acknowledge stimulating discussions with S.-J. Oh, E.-J Cho, and J. Allen, and thank F. Assaad, J. Werner, Bohm Jung Yang, and J. Kroha for helpful discussions. This research was supported by the National Research Foundation of Korea (NRF) grant funded by the Korean government (MSIP) (No. 2011-0028736, 2008-0062606, 2013K000315), by the Institute for Basic Science (IBS) in Korea, and by the Deutsche Forschungsgemeinschaft (Grants No. FOR1162) and the Bundesministerium f\"ur Bildung und Forschung (Grant No. 03SF0356B).\\

\section{\textbf{Supplementary information}}
\setcounter{figure}{0}
\makeatletter 
\renewcommand\thefigure{S\arabic{figure}}
\makeatother

\subsection{\textbf{Methods summary}}
High-quality SmB$_6$ single crystals were cleaved {\em in situ}
along the (001) plane at 40\,K by using a cleavage post glued on top of
the sample. The high-resolution ARPES experiments were performed with a VG
Scienta R4000 detector at the UE112-PGM-1b (``$1^{3}$'') beamline of
BESSY~II at the
Helmholtz Zentrum Berlin (HZB), using both horizontally
and circularly polarized light at $h\nu=70$~eV. At this photon
energy the relative photoemission cross-section is in favor of the $4f$ states
as compared
to ARPES experiments in the laboratory with typically $h\nu \lesssim
40$~eV. All data have been measured with horizontally polarized light, if not stated otherwise. The photoelectrons at $E_F$ with
using $h\nu$\,=\,70\,eV represent $k_{z}$\,=\,6$\pi/a$ in normal
emission using the inner potential 13.5\,eV 
($a$ is the lattice constant \cite{sirota_temperature_1998}),
which means the constant energy sphere covers the
$\Gamma$--$X$--$M$ plane of the cubic BZ in normal emission
\cite{miyazaki_momentum-dependent_2012}. The total energy resolutions
were 7\,meV and 10\,meV for ARPES maps and FS mapping,
respectively. The ARPES maps were measured outside the first
Brillouin zone because of the higher intensity due to 
matrix element effects. The normalized ARPES data
are achieved by taking a logarithmic scale for each
MDC$_{E_B}(k)$ divided by the minimum of the
MDC$_{E_B}(k_{min})$ in the ARPES intensity map.\\

\subsection{Fig. S1}
\begin{figure}
	\centering
			\includegraphics[width=1.0\textwidth]{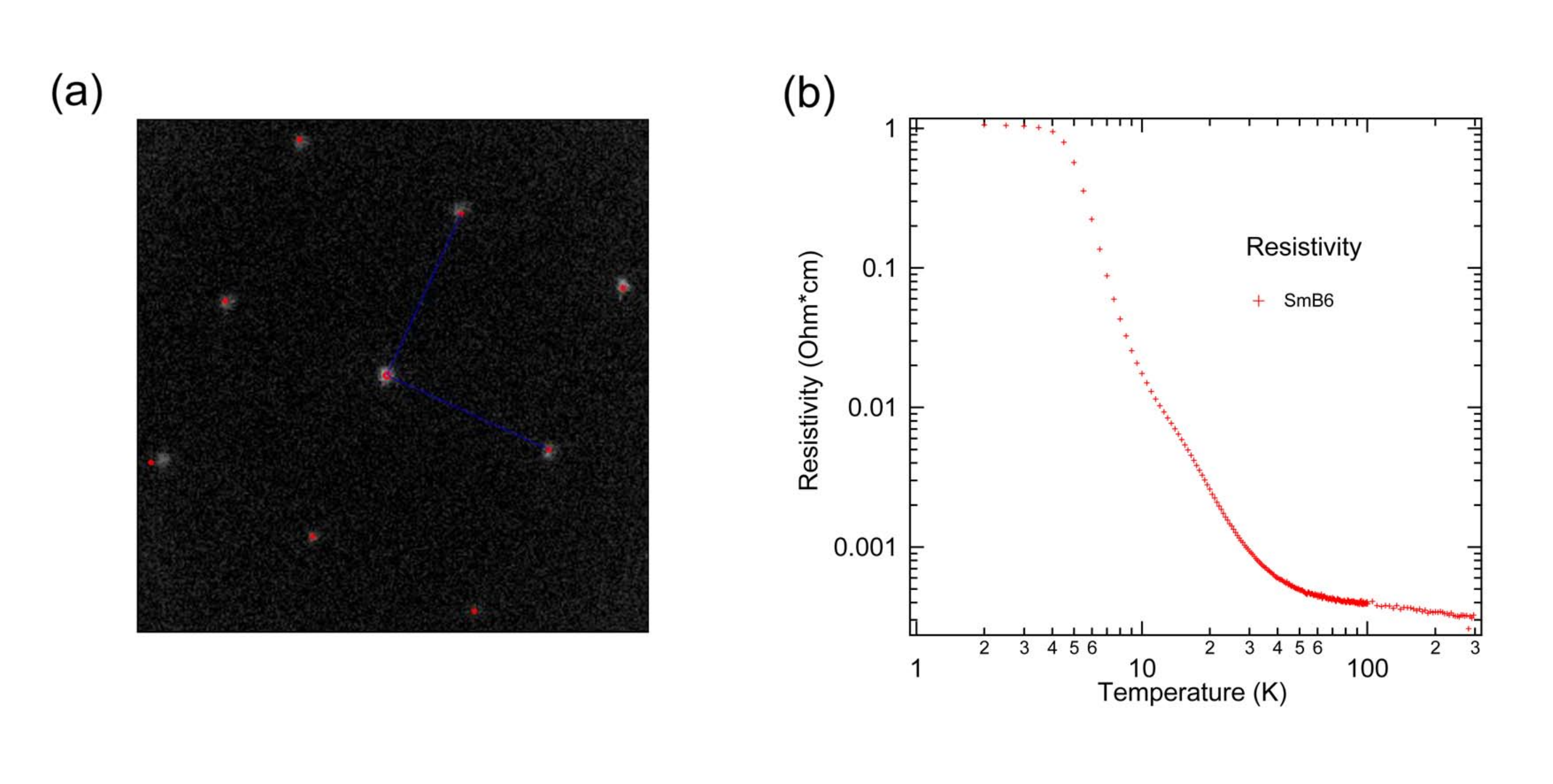}
	\caption{\textbf{SPA-LEED pattern of the cleaved surface and resistivity.} (a) The SPA-LEED pattern of SmB$_6$ single crystal's (100) surface with an electron energy of  99.6\,eV. It was cleaved below 100\,K. The 1$\times$1 diffraction pattern is compared with simulation results using a lattice constant 4.13\,\AA~(red dots). (b) The resistivity suddenly rises with decreasing temperature at 40\,K that can be explained with our ARPES measurements.}
	\label{fig:figS1}
\end{figure}
\indent We measured the Spot Profile Analysis Low-Energy Electron Diffraction (SPA-LEED) of cleaved single crystals along the (001) direction below 1.0$\times$10$^{-10}$\,mbar (Fig.\,\ref{fig:figS1}\,(a)). The crystals were cleaved below 100\,K as we performed in the ARPES measurements. The SPA-LEED showed a 1$\times$1 diffraction pattern. This pattern was reproduced by more than 5 different cleavages at different temperatures, and there was no change in diffraction pattern by varying the temperature from 100\,K to 300\,K in our SPA-LEED resolution.\\
\indent The resistivity of SmB$_6$ single crystals are measured with a four point-contact method (Fig.\,\ref{fig:figS1}\,(b)). The sudden increase in resistivity around $T$\,=\,40\,K can be explained with our ARPES data, which can be understood as consequence of the E$_F$ shift. E$_F$ is inside the gap at $T$\,$=$\,40\,K.\\
\subsection{Fig. S2}
\begin{figure}
	\centering
				\includegraphics[width=1.05\textwidth]{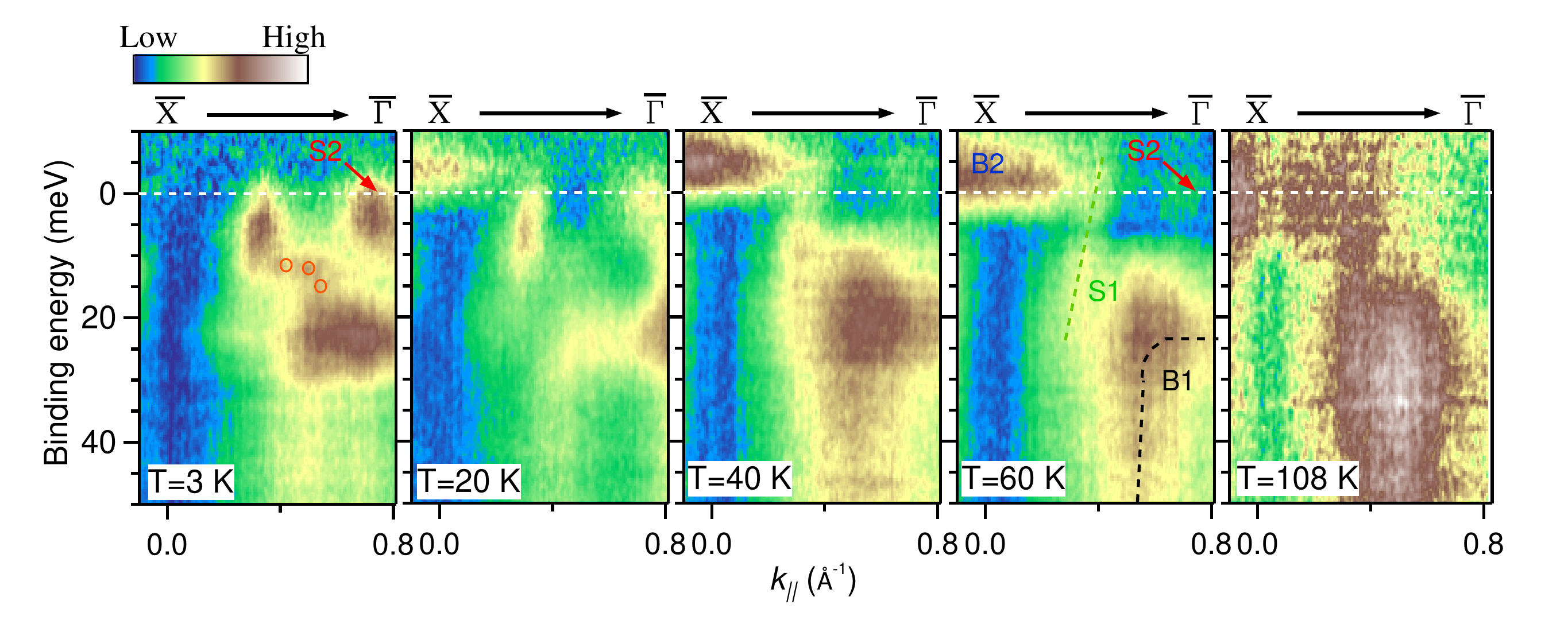}
	\caption{\textbf{Hybridization between $d$\,band and
           $4f$ state ($^{6}H_{5/2}$), development of surface states.}
          ARPES data from Fig.\,2\,(a)  on a logarithmic intensity scale. In the map at $T=60$~K, B1, B2, S1, and
          S2 are indicated by lines and arrows of their specific colors. Remarkably, even at $T=108$\,K, B2 is
          distinguishable from B1 because the
          hybridization makes the emission intensity of $^{6}H_{5/2}$ 
          discontinuous in $k$. At $60$~K, $E_F$ is at the bottom of
          B2. At 3\,K, the states appearing around  $k_\|=0.5$~{\AA} (marked by orange circles) is
          clearly developed, and only faintly visible at
          20~K.} 
	\label{fig:fig3}
\end{figure}

In order to separate dispersive bands from $k$-independent features
($e.g.$ peak broadening due to many-body scattering
\cite{martin_theory_1979}, spectral weight of the non-dispersive
$f$ multiplet, etc.), the ARPES intensity in Fig.\,\ref{fig:fig3} is
plotted on a normalized logarithmic scale. S1 \cite{xu_surface_2013} appearing at all temperatures could be a surface resonance because the area of Fermi surface sheet in Fig.\,1\,(a) changes as a function of temperature. It mainly disperses like $5d$\,bands and smears out where the projected bulk bands are located. It may be less hybridized with the bulk $^{6}H_{5/2}$ feature than the bulk $d$\,band. At $T$\,=\,108\,K, B1 can be distinguished from B2 because the $^{6}H_{5/2}$\ state hybridizes with the $d$\,band so that its intensity drops around the $\bar{\text{X}}$\,point. However, there is a strong thermal broadening on the given small energy scale, mainly because of electron-electron and electron-phonon scattering, which leads to
spectral intensity inside the gap region. Whereas at high temperatures ($cf.$ $T$\,=\,108\,K) $E_F$ is just crossing the bulk band B2, at lower temperatures, B2 shifts upwards until $E_F$
lies below B2 at $T\leq40$\,K, i.e. inside the gap. Therefore, the gap evolution of SmB$_6$ deviates
from that of a conventional KI whose gap starts to open right at
$E_F$. 
At $T\leq20$\,K, the bands in the ARPES data become
clearer because of the reduced thermal broadening. In particular,
there appears a new band around $k_\|=0.5$~{\AA} (marked by orange
circles) which is not discernible at higher temperatures. The well
developed bands and the observed hybridization gaps are a clear
indication that the system is below the coherence temperature $T^{*}$ \cite{hewson__1993}.

\subsection{Fig. S3}
\begin{figure}
	\centering
			\includegraphics[width=0.55\linewidth]{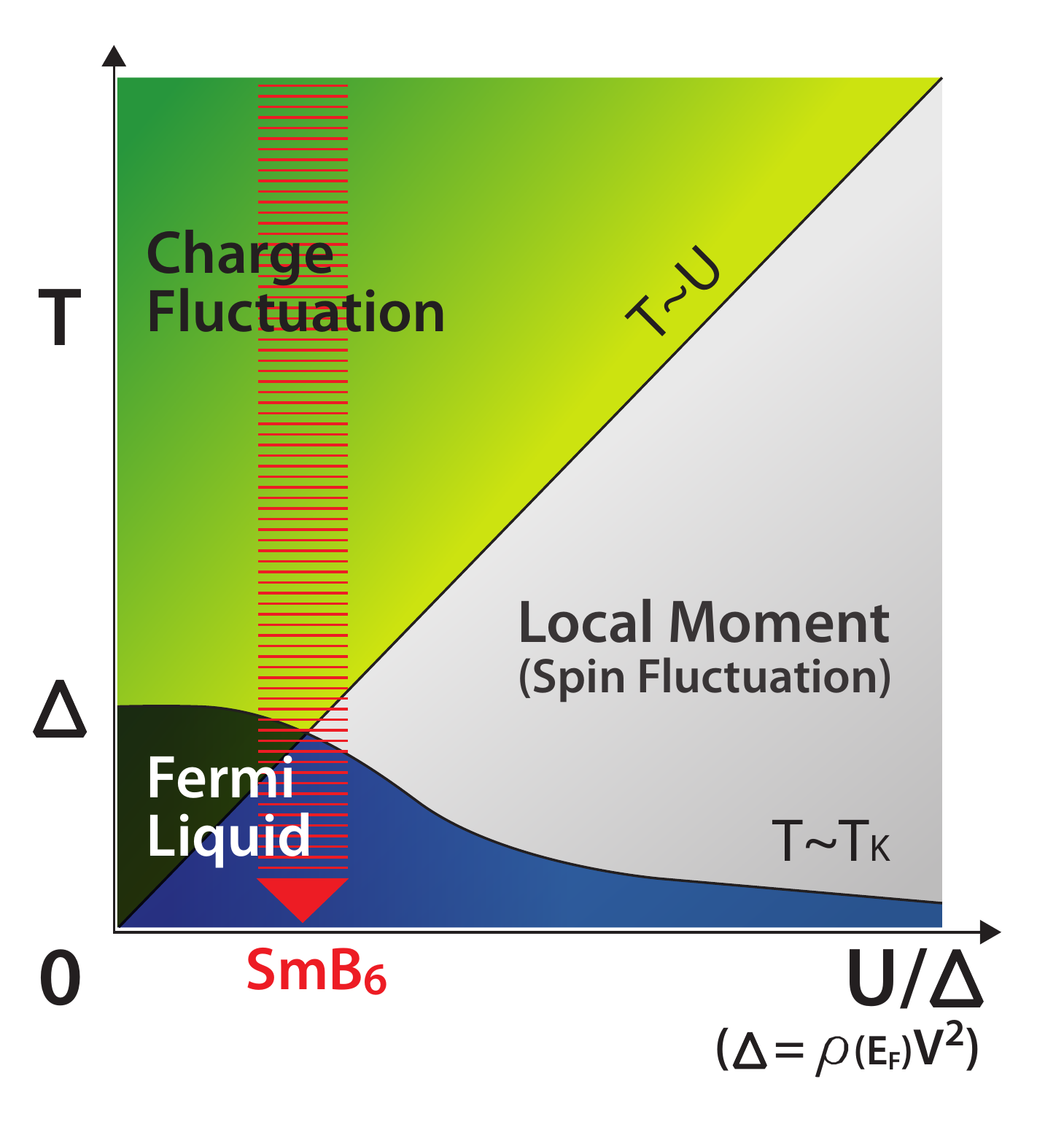}
	\caption{\textbf{Phase diagram of the symmetric Anderson
            model \cite{tanaskovic_phase_2011, otsuki_evolution_2009,
              coleman_introduction_2011}.} The variation of the model
          parameters U/$T$/$\Delta$ (electron-electron coulomb interaction in impurity site/temperature/ A function of hybridization and density of states at $E_F$) separate three different phases:  charge (valence)
          fluctuation on the top left side of the diagonal defined by $T=U$, local moment (spin fluctuation)
          regime on the bottom-right. In the local moment regime at below Kondo temperature $T_K$, the local moment is screened by conduction electrons and becomes a local Fermi liquid in the single impurity case.} 
\label{fig:figS3}
\end{figure}

 In the single Kondo impurity case, the local moment is screened by conduction electrons and becomes a local Fermi liquid below the Kondo temperature. In the lattice case, the coherence effects among $f$\,electrons become dominant, and the coherence temperature $T^{*}$ could be the relevant parameter to describe the Fermi liquid phase ($T^{*}$ is usually considered to be lower than $T_K$). SmB$_6$ has been considered as a Kondo insulator which is at high temperatures in the local moment regime of the Anderson model, and in the Fermi liquid regime at low temperatures \cite{tanaskovic_phase_2011, otsuki_evolution_2009,coleman_introduction_2011}. Local moments $i.e.$ spin fluctuations play a major role in the low temperature physical properties. However, the relevant regime for SmB$_6$ is at the left part of the diagram in Fig.\,\ref{fig:figS3} with the red dashed area that represents  the mixed valence regime. Here, $U/\Delta$ is smaller than in the local moment regime \cite{hewson__1993}. $T^{*}$ is higher in the mixed valence regime than in the local moment regime so that both spin and charge fluctuations play an important role in this Fermi liquid regime.

\subsection{Fig. S4}
\begin{figure}
	\centering
			\includegraphics[width=1.01\linewidth]{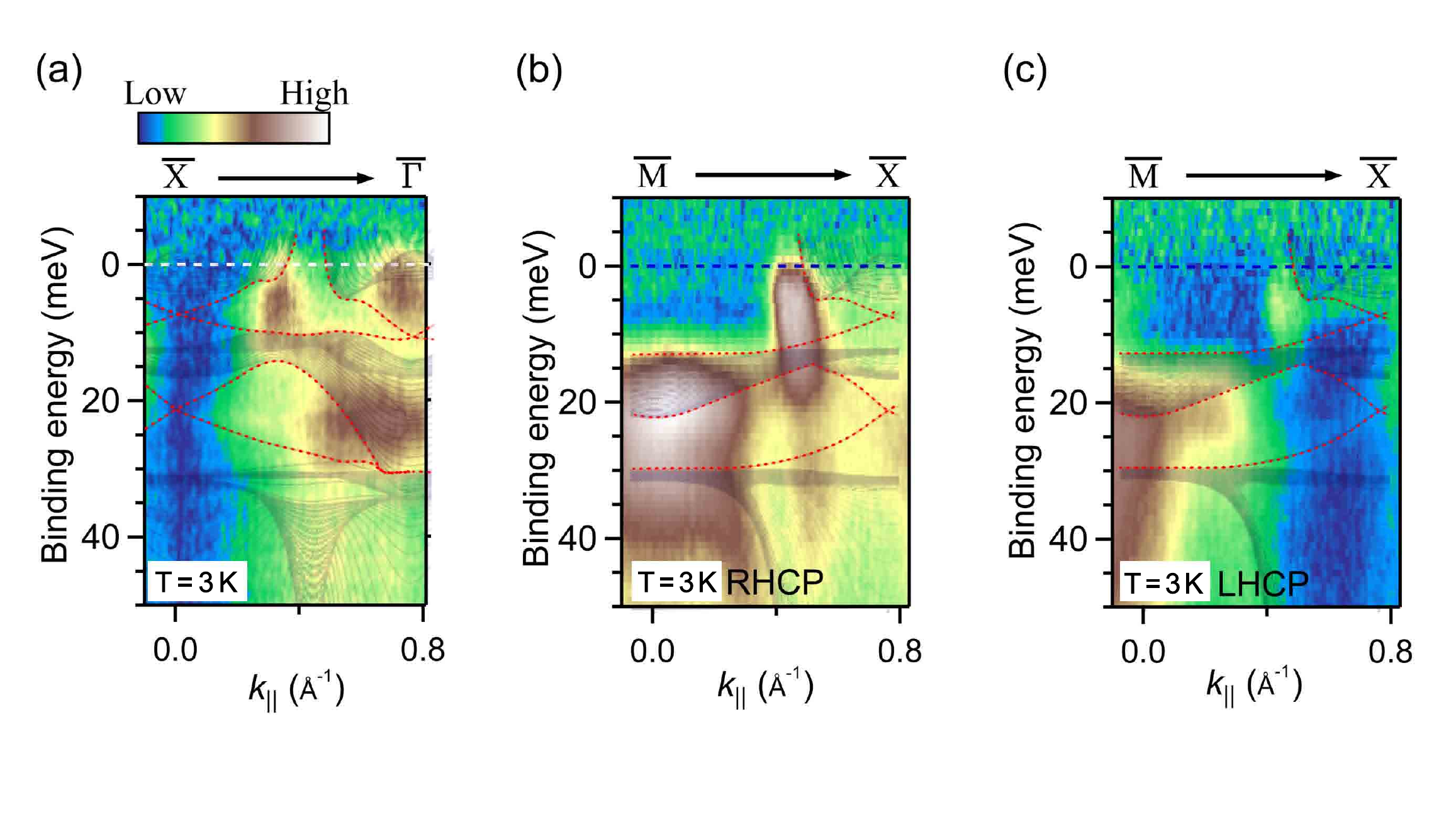}
	\caption{\textbf{Calculated topological surface states at the
            $\bar{\text{X}}$ point.} Theoretically predicted band
          structures \cite{lu_correlated_2013} were overlapped
          transparently onto ARPES
          data at $T=3$\,K along (ii) and (i). The energy scale
          of the calculated band structures was reduced to fit with the
          experiment. The red dotted lines represent the predicted topological
          surface states (TSS). The ARPES data in (a), (b), and (c) are
          obtained along (ii), (i), and (i) in Fig.\,1\,(a),
          respectively, which were measured with horizontally
          polarized, right handed circular polarized (RHCP) and left handed circular polarized (LHCP) light,
          respectively. The data  at $T$\,=\,3\,K show an enhanced
          intensity where the TSS is predicted. At $\bar{\text{X}}$
          and E$_B$\,=\,$20$\,meV, we observe further bands dispersing
          similar to the calculated TSS in Fig.\,4.} 
\label{fig:figS4}
\end{figure}

The calculated band structures from Lu \textit{et al.}
\cite{lu_correlated_2013} are superimposed on the experimental ARPES data at
$T$\,=\,3\,K, shown already in Fig.\,4. We can compare the
$^{6}H_{5/2}$ feature and the theoretical calculation because the
$^{6}H_{5/2}$ multiplet \cite{martin_theory_1979} has the same
symmetry as the ground state configuration for n$_{f}$\,=\,6 in LDA
\cite{alexandrov_cubic_2013, lu_correlated_2013}. The energy
scale from the calculation was compressed about 50\,\%  and
shifted to match the experimental bands around $E_B$$=$20\,meV, because
the $\Gamma^{6}$ states, which are considered in the calculation, do not appear at the same energy position in the ARPES data. Therefore, B1 disperses until it overlaps with
the next $4f$ multiplet feature at $E_B\approx30$\,meV. 
The red dotted lines and the dark blue lines represent the theoretically predicted TSS and projected bulk
band structures, respectively. As shown in Fig.\,2, we could
see that the thermal broadening is considerably reduced at
$T<20$\,K. In comparison with the data at $T=20$\,K,
the intensities at 3\,K become more pronounced in the
region where the TSS is predicted. Furthermore, there appears
some additional intensity at $\bar{\rm X}$ and $E_B=5$\,meV in
Fig.\,\ref{fig:figS4}\,(b)\,\&\,(c), where also TSS is predicted. However, a clear dispersion is not
discernible within the experimental uncertainties although our data
are measured with high energy resolution. The
intensity at the $\bar{\rm X}$ point is only observed in 
$\bar{\rm M}$--$\bar{\rm X}$ direction. Such an elliptical FS, which has its longer axis along
$\bar{\rm M}$--$\bar{\rm X}$,  being
due to a TSS at $\bar{\rm X}$ point has been predicted from recent
theoretical work \cite{alexandrov_cubic_2013}.

Further investigations are necessary to identify the correct
number of $f$ and $d$\,bands. Since the ground state configuration of the $f$ and $d$ states in
the cubic symmetry is not well identified yet
\cite{alexandrov_cubic_2013, lu_correlated_2013}, the number of $f$
bands should be clearly determined because this number will change the sign of the products of parity of occupied states at the time reversal invariant momenta (TRIM) points in Z$_2$ topology
\cite{fu_topological_2007-1, fu_topological_2007}. 
\end{document}